\documentclass[twocolumn,prl,aps,showpacs,superscriptaddress,amsmath,amssymb]{revtex4}

\usepackage{graphicx}
\usepackage{caption}
\usepackage{pstricks}
\usepackage[dvips]{hyperref}
\usepackage[TS1,OT1,T1]{fontenc}
\newrgbcolor{mod}{1.00   0.0   0.0}

\begin{document}
\title{Kinetics of heterogeneous nucleation and growth: An approach based on a grain explicit model}
\author{B. Rouet-Leduc}
\affiliation{CEA, DAM, DIF, F-91297 Arpajon, France}
\affiliation{D\'epartement de Chimie, Ecole Normale Sup\'erieure, 24 rue Lhomond, 75231 Paris cedex 05, France}
\author{J.-B. Maillet}
\email{jean-bernard.maillet@cea.fr}
\affiliation{CEA, DAM, DIF, F-91297 Arpajon, France}
\author{C. Denoual}
\affiliation{CEA, DAM, DIF, F-91297 Arpajon, France}
\date{\today}
\begin{abstract}
A model for phase transitions initiated on grain boundaries is proposed and tested against numerical simulations: this approach based on a grain explicit model (GEM) allows to consider the granular structure, yielding accurate predictions for a wide span of nucleation processes. Comparisons are made with classical models of homogeneous (JMAK \cite{JMAK}) as well as heterogeneous (Cahn \cite{Cahn}) nucleation. A transition scale based on material properties is proposed, allowing to discriminate between random and site saturated regimes. Finally, we discuss the relationship between an Avrami type exponent and the transition regime, drawing conditions for its extraction from experiments.
\end{abstract}
\maketitle

Recrystallization is a mechanism of great scientific and technological importance, encountered during the thermomechanical processing of various materials including metals \cite{Christian}. The first model that efficiently captures the main features of crystallization, namely the JMAK model, states that grains nucleate from points of random locations, and that the grains grow until impinging other neighboring growing grains. Thanks to its straightforwardness, this model was also used in many other situations, provided that the hypotheses of random nuclei and independent growing zone are met: temperature dependant crystallization \cite{Farjas}, combustion \cite{Karttunen}, particle physics \cite{Csernai},  crystallization in amorphous materials \cite{Spinella}, evolution of damage under dynamic tensile loadings \cite{Trumel}, and solid state phase transitions in general \cite{Mittemeijer}, making the JMAK model a much encountered approach.

Random distribution of nuclei is however a rather crude hypothesis. To give examples, materials experience damage by crack nucleation preferably at grain boundaries \cite{Kobayashi, Chen}, combustion of solid energetic materials starts at preferred sites \cite{Howe}, nucleation of grains during recrystallization appears at prior grains frontier \cite{Stipp}, microstructuring nanocomposites enhances the kinetics of physisorption \cite{Jeon}, and crystallization can be influenced by impurities \cite{Granasy} or confinement in a porous media \cite{Woo} or contact with grain boundaries of an other material \cite{Wang}. Thus, the problem of nucleation from interfaces is encountered in a variety of fields \cite{Humphreys,Daniels,Bieler,Was,Sankaran,Coffey,Murata} and the importance of structured nucleation sites - inner nucleation free volumes bounded by interfaces - as well as grain size dependence is commonly witnessed \cite{Cahn,Massoni}.

In this regard, extensions of the JMAK model have been proposed over the years to take into account the specificities of structured nucleation. Most derivations still consider random distribution of nuclei, improving only marginally the model by fitting so-called Avrami parameters, which is seen as lacking clear physical justification \cite{Jagle,Starink,Cahncone}. 

One of the major improvements of the JMAK model that faces the problem of nucleation heterogeneity has been proposed by J.W. Cahn and considers nuclei distributed on planar interfaces \cite{cahn55,Cahn}. This approximation extends the predictions to heterogeneous nucleation, provided that the density of nucleation sites remains low enough. 
Cahn's main assumption is that the superimposing planes are randomly located. However, an accurate modeling for the higher nucleation densities requires to capture the deterministic nature of the location of nucleation -the grains boundaries- excluding grain volume as a possible nuclei source. In other words, an assembly of random planes is a coarse description of interfaces in a granular material.

In this Letter, we propose an accurate modeling of the kinetics of grain nucleation and growth that takes explicitly into account grain boundaries as preferred sites for nuclei. A characteristic length $L_t$ is introduced and compared to the average grain size, allowing for a detailed analysis of the influence of the microstructure on the kinetics of transformation. It is shown that for high nucleation rates, the kinetics are controled by the granular structure, leading to a deterministic behavior. On the other hand, for a decreasing nucleation rate, the effect of microstructure progressively vanishes. We predict the transition between these two regimes. The validity of this modeling is supported by simulations of transformations initiated by random nucleation on the interfaces of a Voronoi tessellation. Finally, we revisit the relationship between an Avrami type exponent and the transition from homogeneous to heterogeneous nucleation.


\begin{figure}[h!]
\begin{center}
\includegraphics[width=8cm]{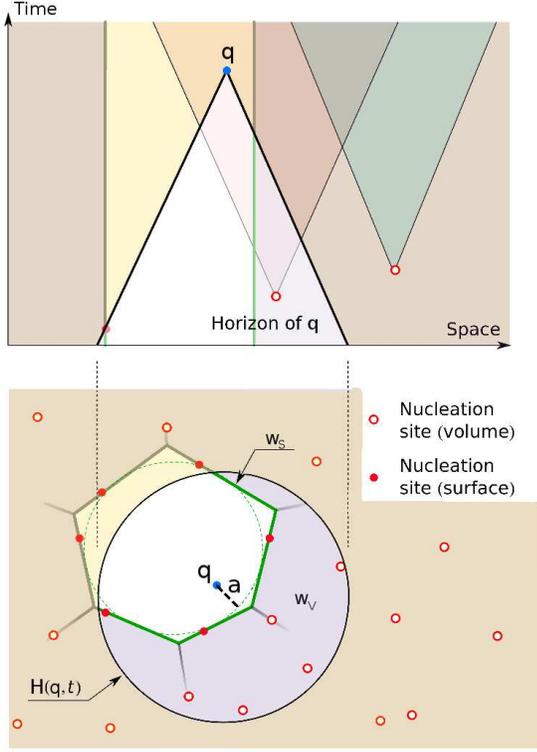}
\end{center}

\caption{{\footnotesize (Color online) Upper picture: Schematic representation of the horizon of a point $q$ in a time-space diagram. Nucleation points are represented as red dots, with their associated (shaded) cone of transformation. The grain is represented by the two vertical lines. Lower picture: Schematic representation of the two sets of nucleation sites within a slice of $H(q,t)$, on interfaces (closed red dots) with intensity $\alpha$ and in volume (open dots) with intensity $\frac{S}{V}\alpha$.}} 

\label{fig:fig1} 
\end{figure}


Once nucleated at time $t'$, a transformation zone expands from the nucleation site over a maximum travel distance given by $h_{t'}(t)=\int_{t'}^t{c(s)ds}$, $c(s)$ being the expansion celerity and $t$ the present time. All potential nucleation sites in this expansion zone then become inhibited. Thus, a nucleation point $q'$ is inhibiting all nucleation in a time-growing transformed zone of radius $h_{t'}(t)$. 


Considering a random point $q$, a nucleation event taking place at time $t'$ at a distance smaller than $h_{t'}(t)$ will transform $q$ before $t$. Therefore the probability that $q$ is transformed at $t$ is given by the probability that at least one nucleation occurred in its horizon $H(q,t)=\{(x,t'),\|x-q\|\le h_{t'}(t)\}$(see Fig. \ref{fig:fig1}-a). 


%


Using the time-cone method (see Ref. \cite{Cahncone} for a demonstration) the transformed volume fraction is expressed as:
\begin{eqnarray}
\Phi(t)=\Phi(q,t)&=&1-\exp\left(-\int_{H(q,t)}{\alpha(q',t') dt' dq'}\right)\nonumber\\
 &=& 1-\exp\left(-N(q,t)\right)
\label{eq:master}
\end{eqnarray}
%

where $\alpha(t)$ is the nucleation rate density, and $N(q,t)$ is therefore the average number of nucleation events over the horizon $H(q,t)$. 

In what follows we give an expression for $N$ in the general case of heterogeneous nucleation, focusing on nucleation on grain boundaries.
Our model is grounded on a simplified representation of a polycrystal as a spherical grain surrounded by an averaged and homogeneous material. 
In this grain explicit model (GEM), nucleation can occur on the grain surface (with intensity $\alpha(t)$) or beyond with an intensity approximated by the average nucleation rate over the whole volume, $\alpha(t)\frac{S}{V}$, as depicted in the lower picture of Fig. \ref{fig:fig1}. 


The spherical symmetry allows to reduce the integration over the grain radius $r$ only. Considering a point $q_a$ at a distance $a$ from the grain boundary, we define the average number of nucleation events $N_r(a,t)$ inside the horizon $H(a,t)$ as a function of $a$ and $r$.


By noting that nucleation is not possible inside the grain, the volume integral in Eq. \ref{eq:master} can be split into two terms, accounting for nucleation at its surface noted $S_r$, and from the outside material noted $\overline{V_r}$: 

%

\begin{equation}
N_r(a,t) = \int_{S_r\cap H(a,t) } {\hspace*{-1.1cm} \alpha (t')dq'dt'} + \frac{S}{V}\int_{{ \overline{V_r}}\cap H(a,t)} {\hspace*{-1.15cm} \alpha (t')dq'dt'}\;, 
\end{equation}  
with $S_r\cap H(a,t)$ the intersection of the horizon with the grain surface and $\overline{V_r}\cap H(a,t)$ the intersection of the horizon with the outside material.
%
The fraction of grain surface $w_S(r,a, \tau)=S_r\cap H(a,t)$ (a spherical cap) is a function of the propagation time $\tau = t-t'$ between a past event occurring at $t'$ and the current time $t$:
\begin{equation}
w_S(r,a, \tau)=\begin{cases} 0 & \text{if $c\tau<a$,}
\\
 \pi\left[c^2\tau^2-a^2\right]\left(\frac{r}{r-a}\right) & \text{if $a \le c\tau\le 2r-a$,}
\\
4\pi r^2 &\text{if $c\tau>2r-a$}\;,
\end{cases} 
\label{ws}
\end{equation}
where $c$ is considered constant for the sake of simplicity. Similarily, the fraction of the averaged volume $w_V(r,a, \tau)=\overline{V_r}\cap H(a,t)$ (a lens) is given by:
%
\begin{equation}
w_V(r,a, \tau)=\begin{cases} 0 & \hspace{-2.4cm} \text{if $b<0$,}
\\
 \frac{4}{3}\pi c^3\tau^3 -\frac{\pi}{3}\left(c\tau+a-b\right)^2 \left(2c\tau-a+b\right) 
\\
\quad - \frac{\pi}{3}b^2(3r-b)
& \hspace{-2.4cm}  \text{if $0\le b\le 2r$,}
\\
\frac{4}{3}\pi c^3\tau^3-\frac{4}{3}\pi r^3 & \hspace{-2.4cm} \text{if $b>2r$\; }\;,
\end{cases} 
\label{wv}
\end{equation}
with $b=\frac{c^2\tau^2-a^2}{2(r-a)}$.
Finally the investigated integral $N_r(a,t)$ becomes a simple time convolution:
\begin{equation}
\hspace{-0.1cm}N_r(a,t) =\hspace{-0.1cm} \int_0^t \hspace{-0.1cm}\left[w_S(r,a, t-t')+w_V(r,a, t-t')\frac{S}{V} \right]\hspace{-0.1cm} \alpha (t')dt'\;.
\label{eq:Phigeneral}
\end{equation}  
We can now express the transformed fraction of a grain $\Phi(t,r)$ as an integration of the probability of transformation $\left[1-\exp\left(- N_r(a,t)\right) \right]$ towards its center following a homothetic path:
\begin{equation}
\Phi(t,r)=\frac{3}{r} \int^r_0{\left[1-\exp\left(-N_r(a,t)\right) \right] \left(\frac{r-a}{r}\right)^2 { \!\!da}}
\label{Phigrain}
\end{equation}  
In what follows, we will show that the presented model is accurate over a wide range of situations, and exact for the limit cases of homogeneous and site saturated nucleation.
%
%

\begin{figure}[h!]
{\includegraphics[trim = 0mm 0mm 0mm 0mm,clip, height=8.6cm,angle=270]{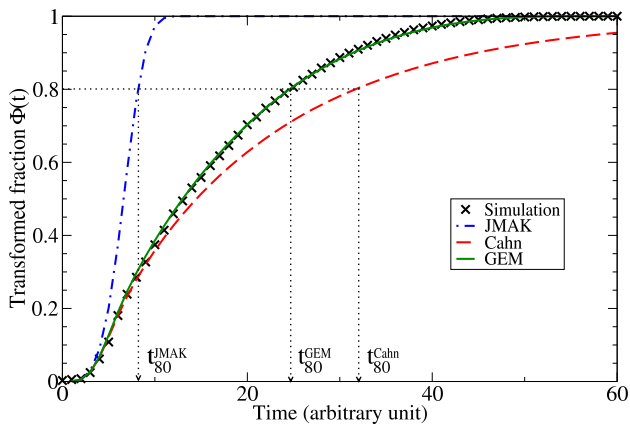}}
\caption{{\footnotesize (Color online) The fraction of transformed material is plotted against the time. Numerical parameters: microstructure generated by a voronoi tessellation of density $5.10^{-6}$, average grain radius $L_g=28.8$ units of length, propagation of the transformation by $c=0.5$ units of length per unit of time, averaged nucleation rate constant and set to $\alpha\frac{S}{V}=0.00275$. Simulations performed on $10^6$ voxels with perdiodic boundary conditions. The Mathematica program computing GEM is supplied as online supplemental material.}}
\label{fig:phivstime} 
\end{figure}

For every low nucleation rate $\alpha(t)$, a significant $N_r(a,t)$ is obtained only for long times $t$ leading to $ct\gg r$ and $w_S=4\pi r^2 \ll w_V = \frac{4\pi}{3} [(t-t')c]^3$ i.e. the transformation of a point is most likely to be caused by a nucleation outside of the grain. The transformed fraction $\Phi(t)$ then reduces to the classical solution of the JMAK model:
\begin{equation}
\Phi(t)=1-\exp\left(-\frac{4\pi}{3} \int_0^t \alpha(t') [c(t-t')]^3 \mathrm{d}t'\right)
\end{equation} 
On the other hand, a nucleation rate high enough ensures that $[1-\exp(-N_r(a,t))]= 1$ for $ct \ge a$ (and $0$ otherwise), yielding the exact expression of the homothetic transformation of a grain, once injected in Eq.\ref{Phigrain}: 
\vspace*{-0.2cm}
\begin{equation}
\Phi(t,r)= \frac{3}{r}\int^{ct}_0  \left (\frac{r-a}{r}\right)^2 da= 1-\left(\frac{r-ct}{r}\right)^3\;,
\label{Homothetic}
\end{equation} 
with $ct\le r$.
On Fig. \ref{fig:phivstime} model predictions are presented against grid based simulation results with Voronoi tessellations accounting for the granular microstructure, nucleation events being randomly generated on grain surface voxels, with a spherical expansion of transformed zones from nucleation sites. The very good agreement of the GEM model with the voronoi based simulations validates the spherical grain approximation as well as the representation of the heterogeneous material as a surrounding equivalent media.
As the JMAK model contains no structural information about the nucleation sites, differences with structured models (Cahn and GEM) become significant, even from very early times. Up to intermediate times the absence of grain-size scale correlations between nucleation sites makes the random planes (Cahn) and grain explicit (GEM) models indiscernable. However, approaching the complete transformation, the increasing difference between the Cahn and GEM models highlights the importance of an explicit grain description.

\emph{Transition scale}
We will now show that the transition between homogeneous nucleation (from the 3D averaged volume) and heterogeneous nucleation (from the 2D grain surface) is associated to a characteristic scale. With the aim to distinguish these two extreme behaviors, we will define a characteristic length for dimension $D=2$ and $D=3$.
In this purpose we consider the horizon that contains one nucleation event on average at a characteristic time $t_c$ \cite{Zcharac}: 
\begin{equation}
\int_0^{t_c} \alpha_D(t_c-t') kc^D(t_c-t')^D \mathrm{d} t' = 1\;,
\end{equation}
with $k$ a shape parameter ($k = \pi$ in 2D and $k=\frac{4 \pi}{3}$ in 3D). For a constant $\alpha$, $t_c$ is given by:
\begin{equation}
t_c = \left(\frac{D+1}{\alpha_D k c^D} \right)^\frac{1}{D+1}\;,
\end{equation}
with  $\alpha_2 = \alpha$ and $\alpha_3 = \alpha S/V$. The radius of the horizon at $t_c$ defines the characteristic length $L_D$ representing half of the average distance between nucleation sites:
\begin{equation}
L_D = \left( \frac{c(D+1)}{k\alpha_D} \right)^\frac{1}{D+1}\;.
\end{equation}
For $D=2$, $L_2$ should be orders of magnitude smaller than
the grain radius $L_g$ to guarantee the 2D nucleation
hypothesis. On the other hand, for $D=3$, homogeneous nucleation can
only hold when $L_3$ is compatible with an averaging over
numerous grains, that is for $L_3 \gg L_g$. A
simple transition definition is to consider the frontier between the
two domains, at $L_2  = L_3 = L_t$, which leads to:
\begin{equation}
L_t=\left( \frac{3c}{\pi\alpha} \right)^\frac{1}{3}\;, 
\end{equation}
Nucleation can thus be defined as homogeneous or heterogeneous
depending on how $L_t$ compares to the average volume to surface ratio $V/S$.  
If $L_t$ is larger than $V/S$, the horizon containing one event covers many grains, thus making the dynamics of nucleation homogeneous. On the other hand if $L_t$ is smaller than $V/S$, the horizon containing one event is smaller than the grain, and the kinetics are that of a heterogeneous nucleation.

\begin{figure}[h!]
{\includegraphics[trim = 0mm 0mm 0mm 0mm,clip, height=8.6cm,angle=270]{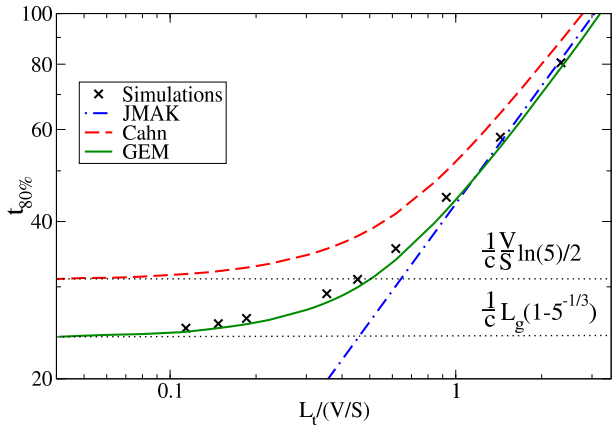}}
\caption{{\footnotesize (Color online) The time to $80\%$ transformation is plotted against the proposed scale $L_t/(V/S)$. The numerical parameters are the same as those of Fig. \ref{fig:phivstime}, except for the nucleation rate $\alpha$, that varies to cover a wide range of nucleation regimes.}}
\label{fig:t80}
\end{figure}

\emph{Discussion} 
The transition from homogeneous nucleation to site saturation is investigated through the use of a wide range of nucleation rates $\alpha$. Recalling that the Cahn and GEM models diverge when approaching complete transformation, we arbitrarily choose the time to $80\%$ transformation $t_{80}$ as a criterium for subsequent evaluation, and compare in Fig. \ref{fig:t80} the various models and the simulation across the transition.
As anticipated, the GEM and Cahn models exhibit a JMAK asymptotic behavior for low $\alpha$ ($L_t\gg(V/S)$). For site saturated nucleation ($L_t\ll(V/S)$) the GEM $t_{80}$ becomes proportionnal to the grain size $L_g$. For site saturated nucleation, the Cahn model exhibits a horizontal asymptote as well. However its limit is different from that of the simulations, and does not explicitely depends on the grain size, but on the volume to surface ratio.
The evolution between these two asymptotic behaviors, reproduced by numerical simulations, evidences the transition occuring over approximately one decade of $L_t/(V/S)$. Since the transition takes place around $L_t=V/S$, the initial guess of a transition scale as $L_2=L_3$ is confirmed to be relevant.

\begin{figure}[h!]
{\includegraphics[trim = 0mm 0mm 0mm 0mm,clip, height=8.6cm,angle=270]{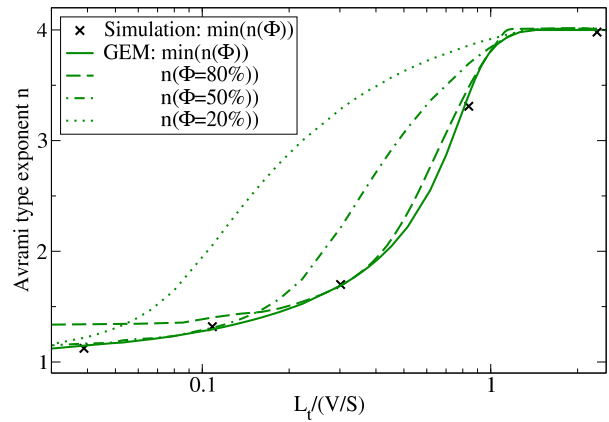}}
\caption{{\footnotesize (Color online) Avrami type exponent plotted against $L_t/(V/S)$. The presented model allows retrieving the avrami exponent from homogeneous ($L_t/(V/S)>1$) to heterogeneous ($L_t/(V/S)<1$) and even site saturated situations ($L_t/(V/S)\rightarrow 0$) using physical parameters: nucleation rate, growth rate, and initial grain size. }}
\label{fig:avrami}
\end{figure}

In addition, a criterium frequently proposed to characterize this transition is:
\begin{equation}
n=\mathrm{d}A(t)/\mathrm{d}\mathrm{ln}(t)\;,
\label{eq:n} 
\end{equation}
with $A(t)=\mathrm{ln}(-\mathrm{ln}(1-\Phi(t)))$, often referred to as the Avrami exponent. This exponent varies from $4$ in case of $3D$ homogeneous nucleation to $1$ in the limit of site saturated grain boundary nucleation. Therefore, it is considered to be a reliable signature of the nucleation regime (heterogeneous vs. homogeneous).
With the aim to provide an unbiased method to determine the Avrami exponent, we propose to define it as the minimum of $n(\Phi)$ (Eq. \ref{eq:n}) and compare it to the commonly used $n$ at fixed transformed fractions. 
On Fig. \ref{fig:avrami}, $\mathrm{min}(n(\Phi))$ (extracted from the GEM model) goes from $1$ to $4$, most of its variation taking place across the previsouly observed transition ($0.1<L_t/(V/S)<1$). Surprisingly, none of the $n$ determined at fixed transformed fraction is able to reproduce both the transition at $L_t/(V/S)\approx1$ and the asymptotic value of $n=1$. Hence, $\mathrm{min}(n(\Phi))$ is the only definition of the Avrami exponent that carries reliable information about the transition between homogeneous and heterogeneous nucleation.

In conclusion, we have proposed a grain explicit model (GEM) for the kinetics of phase
transformation initiated at interfaces, that reconciles heterogeneous and homogeneous nucleation. 
The GEM model exhibits exact limits (JMAK and site saturated), and is validated against numerical simulations spanning all nucleation regimes.
Furthermore we proposed a reliable transition scale $L_t$, based on material properties, which enables the prediction of the nucleation regime once compared to the characteristic length $V/S$ of the granular structure. 
Finally we revisited the determination method of the Avrami type exponent, commonly derived from experimental data, and showed that once defined as the minimum slope of avrami type plots, it offers a second and independant way to retrieve information about the nucleation regime.



\begin{acknowledgments}
The authors would like to thank N. Desbiens and C. Matignon for their useful point of view. D. Hassine is also thanked for his helpful review.
\end{acknowledgments}




\appendix

\end{document}